# Sub 100-ps dynamics of the anomalous Hall effect at THz frequencies

T. J. Huisman[1], R. V. Mikhaylovskiy[1], A. Tsukamoto[2], L. Ma[3], W. J. Fan[3], S. M. Zhou[3], Th. Rasing[1] and A. V. Kimel[1]

[1]*Radboud University Nijmegen, Institute for Molecules and Materials, 6525 AJ Nijmegen, The Netherlands*
[2]*College of Science and Technology, Nihon University, 7-24-1 Funabashi, Chiba,* Japan
[3]*Shanghai Key Laboratory of Special Artificial Microstructure and Pohl Institute of Solid State Physics and School of Physics Science and Engineering, Tongji University, Shanghai 200092, China*

**We report about the anomalous Hall effect in 4f 3d metallic alloys measured using terahertz time-domain spectroscopy. The strength of the observed terahertz spin-dependent transport phenomenon is in good agreement with expectations based on electronic transport measurements. Employing this effect, we succeeded to reveal ultrafast dynamics of the anomalous Hall effect which accompanies the sub-100 picosecond optically induced magnetization reversal in a GdFeCo alloy. The experiments demonstrate the ability to control currents at terahertz frequencies in spintronic devices magnetically and ultrafast.**

To develop ultrafast spintronics logic protocols for high-speed data processing, one has to find a way to control and detect the spin-dependent transport ultrafast. The conventional approach to study magnetotransport is based on the application of an electrical voltage to a sample and the detection of the subsequent electrical current via electrodes. However, making electrical contacts that transfer a picosecond (ps) short signal without substantial distortions and broadening is rather challenging [1]. An elegant solution for the problem of ultrafast conductivity measurements can be found by using freely propagating terahertz (THz) waves.

To date, THz spectroscopy has been used to study magnetotransport in a large diversity of materials [2-17]. However, only a few of these studies have been dedicated to magnetically ordered materials and in none of these materials the magnetization could be reversed ultrafast. To reveal sub-100 ps dynamics in the anomalous Hall effect, we have chosen ferrimagnetic $Gd_x(FeCo)_{1-x}$ thin films, as they are expected to show a significant anomalous Hall effect [18-20], and are known for their ultrafast all-optical magnetization reversal [21-26].

The studied $Gd_x(Fe_{0.875}Co_{0.125})_{1-x}$ samples are 20 nm thick amorphous films, which have adjacent 5 nm SiN protecting layers and are deposited using magnetron sputtering. These samples have a ferrimagnetic ordering and an out-of-plane magnetic anisotropy in the studied Gd concentration range (0.2<x<0.3), with the magnetization being dominated by the FeCo magnetic sublattice at room temperature. At lower temperatures the magnetization of the Gd magnetic sublattice may start to dominate. The temperature at which this happens is known as the magnetization compensation temperature, the value of which can be tuned with the concentration of Gd.

Our experimental setup consists of a THz time domain spectrometer designed to allow simultaneous excitation of the sample with an intense optical pump and detecting changes in a

THz probe pulse, see also Fig. 1(a). The polarization of the incident THz radiation is set by a wiregrid polarizer. Another pair of wiregrid polarizers put after the sample is used to reconstruct the polarization state of the THz radiation transmitted through the sample. The sample is situated inside a cryostat with a superconducting magnet and the bias magnetic field is applied along the normal of the sample. In section 1 of the supplementary information we discuss our experimental setup in more details.

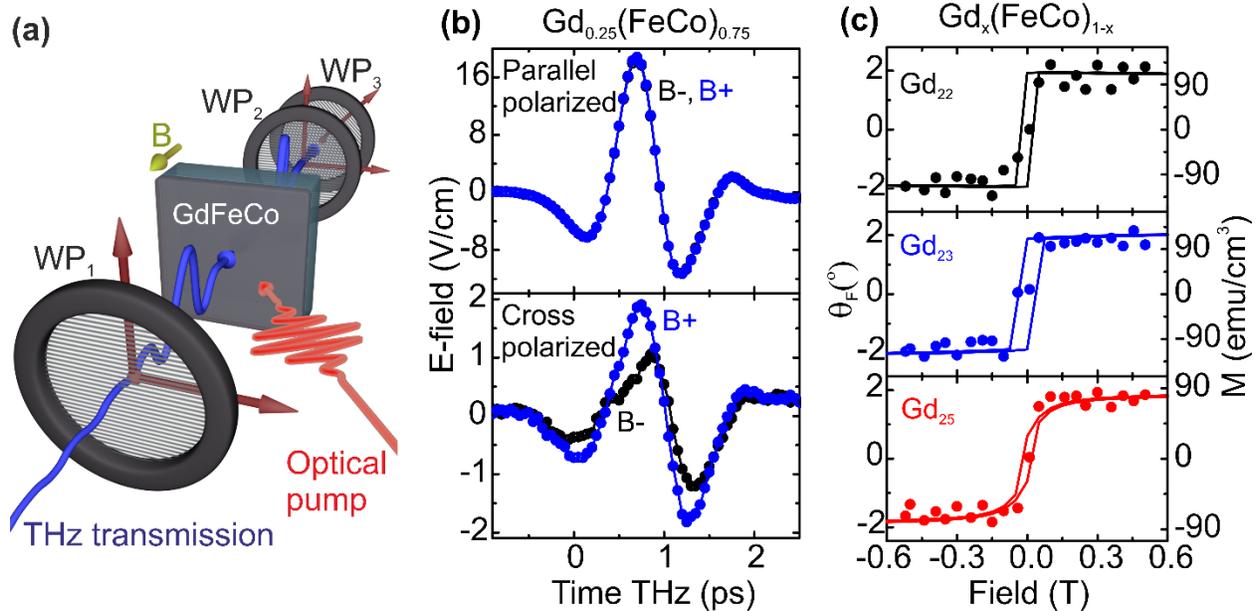

Figure 1: Schematics and room temperature measurements. (a) THz radiation is transmitted through GdFeCo films. The wiregrid polarizers WP$_1$, WP$_2$ and WP$_3$ allow for polarization resolved measurements. Additionally an optical pump pulse can be used to change the transport properties of the metallic films. (b) Observed THz transmission traces for the Gd$_{0.25}$(FeCo)$_{0.75}$ sample. The top panel shows the transmission when WP$_1$ and WP$_2$ are parallel to each other, while the lower panel shows the transmission when WP$_1$ and WP$_2$ are 90$^0$ rotated with respect to each other. B+ corresponds to an applied field of 0.3 Tesla, and B- corresponds to an applied field of -0.3 Tesla. (c) The dots represent the Faraday rotation at THz frequencies for the Gd$_x$(FeCo)$_{1-x}$ samples, shown as function of applied field and averaged between 0.5 and 1 THz. The solid lines are measurements of the magnetization using a vibrating sample magnetometer.

In Fig. 1(b) we show examples of the transmitted terahertz waveforms, measured without the laser pump present, for the Gd$_{0.25}$(FeCo)$_{0.75}$ sample. The top window of this figure shows the transmission when the THz polarizers before and after the sample are oriented parallel. It is clear that the transmitted electric field is time resolved with a sub-picosecond resolution and that applying an external magnetic field of opposite polarity does not induce significant changes to the observed electric field. The lower panel of Fig. 1(b) shows the electric field for the case when the THz polarizers before and after the sample are crossed. The difference between the outcomes of the experiments with parallel and crossed polarizers reveals a rotation of the polarization of the THz radiation. Both magnetization dependent and magnetization independent polarization rotations are observed here, the former is ascribed to the Faraday effect in the sample, while the latter arises due to linear dichroism in the windows of the cryostat.

Experimentally we can determine the Faraday rotation $\theta_F$ by determining the absolute rotation as described in [27], for opposing applied magnetic fields strong enough to saturate the magnetization, taking the difference and dividing by two. Similarly we can do so for the magnetization induced ellipticity ($\eta$). These measurable quantities are expected to be related to the electronic transport properties in thin conducting metallic films according to:

$$\theta_F \approx \frac{\sigma_{xy}(0)}{\sigma_{xx}(0)} \frac{1}{1+(\omega\tau)^2}, \qquad (1)$$

$$\eta \approx \frac{\theta_F \omega \tau}{2}, \qquad (2)$$

where $\sigma_{xy}(0)$ is the off-diagonal or anomalous Hall effect conductivity at zero frequency, $\sigma_{xx}(0)$ is the longitudinal conductivity at zero frequency, $\omega$ is the angular frequency and $\tau$ is the electron scattering time. For a full derivation of Eqs. (1) and (2), we refer to section 2 of the supplementary information. For GdCo and GdFe thin films with a Gd concentration close to 25%, literature reports $\frac{\sigma_{xy}(0)}{\sigma_{xx}(0)}$ to be in the range from 0.02 to 0.04 [18-20]. Since for metals a scattering time in the order of 10 femtoseconds is common [28], we can expect to measure a Faraday rotation in the order of 1 or 2 degrees and a negligible ellipticity. As we will show next, this is exactly what we retrieved experimentally.

In Fig. 1(c) we show with dots the measured Faraday rotation at THz frequencies as function of magnetic field for three different $Gd_x(FeCo)_{1-x}$ samples, without a pump present. Every dot represents the mean THz rotation averaged between 0.5 and 1 THz, measured for two orthogonal incident polarizations. In section 3 of the supplementary information we show similar measurement preformed for different $L1_0$ $FePd_{1-x}Pt_x$ films. Only the $Gd_{0.2}(FeCo)_{0.8}$ sample showed a small but measurable in-plane anisotropy and a magnetic field induced ellipticity. The Faraday rotations of Fig. 1(c) are compared with the solid lines representing the magnetization as measured with a vibrating sample magnetometer (VSM). An appreciable agreement can be found between the two different data sets.

It is known that both magnetotransport and the Faraday effect can show magnetic sublattice sensitivity [29-33]. To confirm whether the THz Faraday rotation has also magnetic sublattice sensitivity, we performed temperature resolved measurements. In Fig. 2(a) we show with dots the mean Faraday rotation at THz frequencies as a function of temperature for the $Gd_{0.25}(FeCo)_{0.75}$ sample. The incident polarization of the THz radiation is kept fixed. It is clear that from 250 K to 265 K the Faraday rotation changes sign as expected when one of the magnetic sublattices is dominating, while the THz transmission (not shown) is nearly independent of temperature between 150 and 300K. Note that the used alternating field of +/- 0.5 Tesla is smaller than the coercive fields near the compensation temperature. To resolve the compensation temperature better, we fixed the THz polarizers in cross polarized orientation and scanned the field for different temperatures, resulting in Fig. 2(b). It is clear that between 256 and 260 K the coercive field diverges and the sign of the observed Faraday rotation changes. The seemingly different onset of the compensation temperature as observed in the Faraday rotation and VSM measurements shown in Fig. 2(a), are likely due to the different calibrations of the temperature used in the different experimental setups.

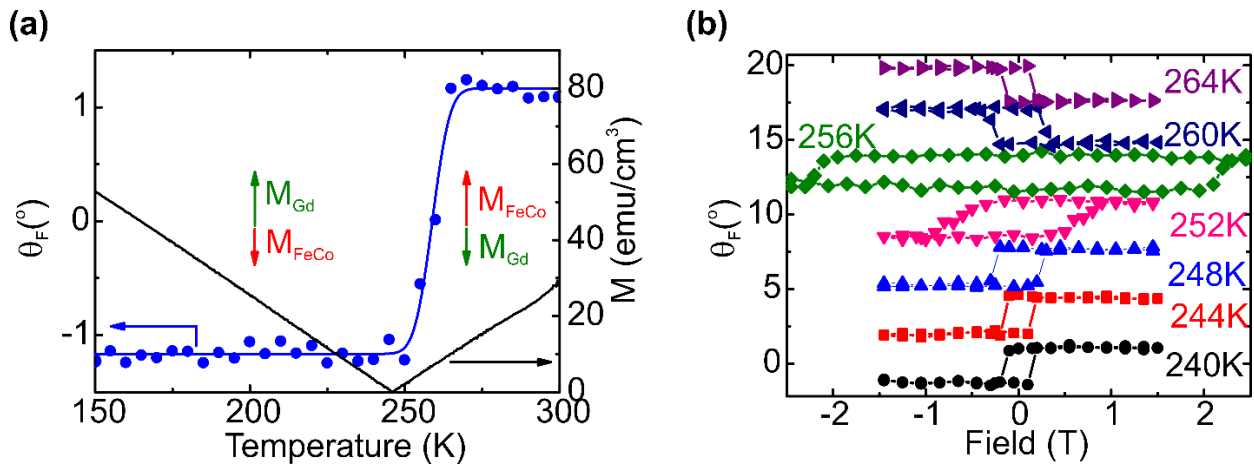

Figure 2: Temperature dependence of the Faraday rotation at THz frequencies for the $Gd_{0.25}(FeCo)_{0.75}$ sample. (a) The Faraday rotation is defined here as the difference in rotation measured for 0.5 and -0.5 Tesla external applied fields, divided by 2 and averaged between 0.5 and 1 THz. The solid black line near the bottom shows the magnetization as measured with a vibrating sample magnetometer, with near zero external applied field. The solid line overlapping the Faraday rotation measurements is a fit used to guide the eye. (b) The peak values of the cross polarized transmission is normalized to give the Faraday rotation as function of field and temperature. The separate graphs are shifted vertically for clarity.

To demonstrate the ultrafast light induced modulation of the anomalous Hall effect, we illuminated the $Gd_{0.25}(FeCo)_{0.75}$ sample with a 50 fs short pump pulse, which has a photon energy of 1.55 eV and a fluence of about 7 mJ/cm$^2$. The top panel of Fig. 3(a) shows the transmitted waveforms of the THz radiation just before the arrival moment of the pump pulse on the sample (closed symbols) and 600 picoseconds after the arrival (open symbols). It is seen from the figure that the femtosecond optical excitation does not change significantly the THz transmission neither for positive, nor for negative applied magnetic field. However, the lower panel of Fig. 3(a) shows that if we perform the same experiment when the incident and detected THz polarization are cross polarized, both the pump and the field induce changes in the observed waveforms. This data shows a clear indication that 600 picoseconds after the arrival time of the pump, a noticeable modification of the Faraday effect is observed, which is most pronounced close to the peak of the THz electric field.

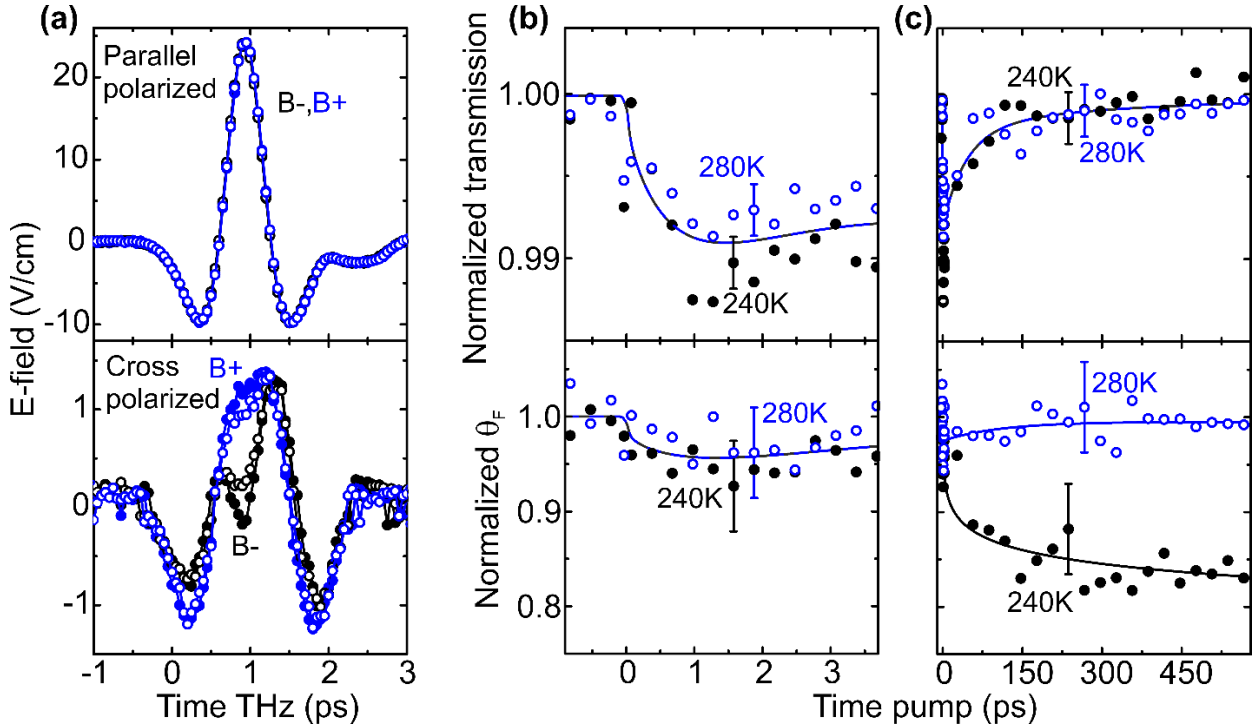

Figure 3: Pump induced dynamics in the transmission of THz radiation in the $Gd_{0.25}(FeCo)_{0.75}$ sample. (a) The transmission is shown when $WP_1$ and $WP_2$ are parallel oriented (top panel) and when they are 90° rotated with respect to each other (lower panel). The blue and black symbols correspond to a 1 and -1 Tesla magnetic field applied, respectively. The closed dots correspond to a pump pulse arriving at the sample 20 ps after the THz radiation past through, while the open dots correspond to a pump pulse arriving at the sample 600 ps before the THz radiation is passing through. The sample temperature is set to 235 K. (b) Pump induced dynamics on the peak of the THz transmission shown in figure (a). The top panel represents the pump-induced dynamics in the normal THz transmission, while the lower panel represents the pump-induced dynamics in the Faraday rotation corresponding to the anomalous Hall effect. The closed black dots correspond to a sample temperature of 240 K, while the open blue dots correspond to a sample temperature of 280 K. (c) Same as (b), but now for a longer timescale. The error bars in (b) and (c) are representative for the estimated standard deviation in these graphs, while the solid lines are fits to guide the eye.

To investigate the origin of the observed photoinduced changes in the Faraday effect we repeated the THz transmission experiment displayed in Fig. 3(a), but now looking at the maximum of the transmitted THz waveform and varying the time delay between the arrival of the pump and THz pulses to the sample. The top panel of Fig. 3(b) shows the photo-induced signal that is even in applied magnetic field as a function of the pump time delay. It is seen that in the first few picoseconds the pump pulses induce a change in the transmission of about 1% which recovers on a time scale close to 100 picoseconds, as shown in the top panel of Fig. 3(c). The lower panels of Fig. 3(b) and (c) show a similar type of experiment, but now for crossed polarizers. Again, we measure the peak of the transmitted terahertz field, scanning the time delay of the pump pulses. We focus on the contribution odd in applied magnetic field that corresponds to the photoinduced change of the Faraday effect. The traces shown in the lower panels of Fig. 3(b) and (c) appear to be very similar to typical pump-probe traces used for monitoring ultrafast magnetization dynamics by means of the Faraday effect, but now with the probe at THz frequencies.

One can see that on short timescales (lower panel of Fig. 3(b)) both below and above the compensation temperature, there is a fast reduction of the Faraday rotation. This ultrafast dynamics of the Faraday rotation can be due to laser-induced dynamics of both electrons and spins. Note that the dynamics of the transmission, which is expected to be dominated by the dynamics of the electrons, is very similar to the dynamics of the Faraday rotation on the timescale of a few ps. However, on the timescale of a few hundred of picoseconds, the Faraday rotation shown in the lower panel of Fig. 3(c) exhibit clearly distinct dynamics compared to the transient transmission, showing that the changes in the Faraday signal are not purely the consequence of a photoinduced change in the conductivity. The character of the Faraday dynamics changes drastically above and below the magnetization compensation temperature, indicating that the Faraday dynamics is affected by the dynamics in the magnetic ordering. As the Faraday signal represents the anomalous Hall effect, we conclude that this observation represents the photoinduced modification of the anomalous Hall effect due to the magnetization dynamics.

Obviously, after laser excitation the magnetization should relax to one of the metastable states. While above the compensation temperature we observe that the Faraday signal is restored to the initial state on a timescale of a few hundred picoseconds, below the compensation temperature but on the very same timescale, it is seen that the system relaxes to another magnetic state. The time dependence of the Faraday signal follows exactly the dynamics expected from laser induced magnetization reversal [22-24,26]. Especially at higher fields, like the ones used here, it can be expected that when the sample is heat-driven across the compensation temperature, the magnetization should reverse. We note that our THz probe has a diameter of about 2.5 mm on the sample, while our pump pulses have a diameter which is almost twice as narrow. It can therefore be expected that a full reversal of the magnetization is not shown as a change in the magnetization of 200%, but much less. Taking also into consideration that most likely only close to the middle of the pump spot the magnetization reverses, a change in the cross polarized THz transmission of a few tens of percent can indeed be expected.

To conclude, we have performed a study on the temporal evolution of terahertz spin-dependent transport with sub-picosecond resolution. Particularly, we have measured the polarization state of the THz-pulse after an interaction with GdFeCo thin films that is connected to the anomalous

Hall effect in these materials. The strength of the anomalous Hall effect at terahertz frequencies can be largely reduced by a femtosecond laser excitation on a sub-100 ps timescale. Our experiment mimics sub-100 ps operation of a spin transistor, in which the source-drain voltage is applied from a freely propagating THz wave and a femtosecond laser pulse plays the role of the gate by acting on the magnetic ordering. Our proof-of-principle experiment demonstrates the feasibility to reveal the fundamental and practical limitations on the speed of magnetic control of conductivity and thus to define the ultimate speed for the operation of spintronic devices.

**Acknowldegments**

We would like to thank T. Toonen, and S. Semin for technical support. This work was supported by the Foundation for Fundamental Research on Matter (FOM), the European Unions Seventh Framework Program (FP7/2007-2013) grant No. 280555 (Go-Fast) and No. 281043 (FemtoSpin), European Research Council grant No. 257280 (Femtomagnetism) and grant No. 339813 (Exchange), the program "Leading Scientist" of the Russian Ministry of Education and Science (14.Z50.31.0034), the State Key Project of Fundamental Research Grant No. 2015CB921501, and the National Science Foundation of China grant No. 51331004 and grant No. 51501131.